\documentclass[proof]{pasj00}

\draft

\begin{document}
\SetRunningHead{Nobuhiro Tanaka}{Suzaku Observations of the Merging Cluster Abell 85}
\Received{}%{yyyy/mm/dd}
\Accepted{}%{yyyy/mm/dd}

\title{Suzaku Observations of the Merging Cluster Abell 85: Temperature Map and Impact Direction}

%%% begin:list of authors
% Do NOT capitalize all letters in "textsc".
\author{Nobuhiro \textsc{Tanaka},\altaffilmark{1}
        Akihiro \textsc{Furuzawa},\altaffilmark{2}
        Shigeru J. \textsc{Miyoshi},\altaffilmark{3}
        Takayuki \textsc{Tamura},\altaffilmark{4}
        and
        Tadafumi \textsc{Takata}\altaffilmark{1}
        }
\altaffiltext{1}{Astronomy Data Center, National Astronomical Observatory of Japan, 2-21-1 Osawa, Mitaka, Tokyo 181-8588}
\email{nobuhiro.tanaka@nao.ac.jp}
\altaffiltext{2}{Department of Physics, Nagoya University, Furo-cho, Chikusa-ku, Nagoya 464-8602}
\altaffiltext{3}{Department of Physics, Kyoto Sangyo University, Kamigamo-Motoyama, Kita-ku, Kyoto 603-8555}
\altaffiltext{4}{Institute of Space and Astronautical Science, Japan Aerospace Exploration Agency, 3-1-1 Yoshinodai, Sagamihara, Kanagawa 229-8510}
%%% end:list of authors

%% `\KeyWords{}' always has to be placed before `\maketitle'.
%Do NOT move this preamble from here!

\maketitle

%%%%%% Abstract %%%%%%%%
\begin{abstract}
%Please read ``IMPORTANT NOTICE'' carefully before preparing a manuscript. 
% 300 words limit
To investigate the present situation of the merging in the southern outer region of Abell 85, 
we carried out long ($\sim$ 100 ks) observations with Suzaku, and produced an X-ray hardness ratio map.
We found a high hardness ratio peak in the east side of a subcluster located in the south of the cluster;
an X-ray spectrum of the region including this peak indicates a high temperature of $\sim$ 8.5 keV. 
This hot spot has not been reported so far.
We consider that this hot spot is a postshock region produced by the infall of the subcluster from the southwest.
By using the Rankine--Hugoniot jump conditions for shocks, 
the Mach number and the infall velocity of the subcluster are obtained as 
$1.5 \pm 0.2$ and $1950^{+290}_{-280}$ km s$^{-1}$, respectively,
in the case of merging with the subcluster from the southwest direction.
By using the redshift difference between the A 85 and the subcluster obtained from optical observations,
the angle between the line of sight and the direction of the motion of the subcluster is estimated to be $75^{+7}_{-8}$ degrees.
We estimate the kinetic energy of the subcluster and the energy used for intracluster medium (ICM) heating
to be $\sim 10^{63}$ and $\lesssim 8 \times 10^{60}$ erg, respectively.
This shows that the deceleration of the subcluster by ICM heating has been negligibly small.
\end{abstract}

%%%%%% 1. Introduction %%%%%%%%
\section{Introduction}
Clusters of galaxies are the largest objects in the universe, and
consist of several tens or more galaxies that are bound together by gravity \citep{Abell}. 
They consist of galaxies, an intracluster medium (ICM) which is a high-temperature plasma gas, and dark matter.
Numerical simulations show that many clusters of galaxies grow by merging
during the formation of large-scale structures (e.g., \cite{Davis}; \cite{Ricker}).
Investigations of merging clusters are necessary not only to understand
the evolution of clusters of galaxies, themselves,
but also to tracing the structure formation of the universe.

Merging clusters that violently come into collision are well studied because
they exhibit interesting features, such as extreme variations in the temperature and the density distributions of ICM
(e.g., \cite{Markevitch2002}; \cite{Markevitch2004}).
However, such intensely merging clusters are very few, and mostly giant cluster cannibalize small clusters or groups of galaxies.
Therefore, it is important to investigate such a general type merging clusters.
Drastic heating of ICM by shock wave is not expected in such cases, and
the early stage of merging occurs in the outer region with low X-ray luminosity.
Therefore, highly accurate temperature determinations and photon-collecting power
are required for investigating of such merging clusters.

As a result of recent advances in X-ray observations, it is possible to find local high-temperature ICM regions
that reveal the existence of merging in galaxy clusters (e.g., \cite{Govoni}; \cite{Bourdin}).
However, even more accurate X-ray observations are required to investigate in detail the shock heating by merging.
The X-ray imaging spectrometer (XIS: \cite{Koyama}) and X-ray telescope on board the X-ray observatory Suzaku have
a low background in the high-energy band and a large collecting area, respectively.
Suzaku is currently the best X-ray observatory for observing general merging clusters
because its instruments are very effective in detecting hot, diffuse regions ,such as shocked regions.

The Abell 85 cluster (hereafter, A 85) is a nearby ($z = 0.055$; \cite{Oegerle})
typical merging cluster in which a giant cluster is cannibalizing small clusters.
A 85 has been observed so far by many X-ray observation satellites
(e.g., ROSAT, \cite{Pislar}; ASCA, \cite{Markevitch1998};
Chandra, \cite{Kempner}; XMM-Newton, Durret et al. 2003, 2005, hereafter D05)
and in other wavelength regions (e.g., \cite{Durret1998a}, \cite{Boue}). 
X-ray observations have revealed a bright subcluster located about \timeform{9'} south of A 85.
Figure 3 in D05 showed that a high-temperature region was found between the main part of A 85 and the subcluster,
and they argued that this ``impact region'' might be thermalized by merging with the subcluster.
However, this map did not cover the wide surrounding regions of the subcluster
because of low signal--to--noise ratio (S/N).

Based on the high statistics temperature map obtained from Suzaku observations,
we investigated the merging situations and the motion of the subcluster.
The Suzaku observations and data reduction are described in \S 2, and
the results of data analysis are given in \S 3.
We discuss for the impact direction, velocity and motion of the subcluster in \S 4.
Finally, the findings of the present study are summarized in \S 5.
Throughout this paper, we assume that $H_0= 70$ km s$^{-1}$ Mpc$^{-1}$,
$\Omega_{\Lambda} = 0.73$, $\Omega_{\mathrm{M}} = 0.27$, and $q_0 = 0.5$.
At the cluster redshift, $z = 0.055$, \timeform{1'} corresponds to 64.2 kpc.
All quoted errors are at the 90$\%$ confidence level unless otherwise stated.

%%%%%% 2. Observations and data reduction %%%%%%%%
\section{Observations and data reduction}
%%%%%% 2.1 Observations %%%%%%%%
\subsection{Observations}
We observed A 85 located at (RA, Dec) $=$ (\timeform{00h41m37s.8}, \timeform{-09D20'33''}) \citep{Abell}
in equatorial coordinates using Suzaku \citep{Mitsuda} on January 2007 with a net exposure time of 99.2 ks.
The Suzaku pointing position is $\sim$ \timeform{8'} south from the center (the peak of X-ray surface brightness distribution) of A 85,
which is (RA, Dec) $=$ (\timeform{00h41m52s.0}, \timeform{-09D25'43''}) in equatorial coordinates.
This position has been termed the ``impact region'' in D05.
The XIS field of view (FOV) is presented at the left side of figure \ref{fig:xmm-img}
on the X-ray image taken by XMM--Newton.

We obtained the XIS and the hard X-ray detector (\cite{Takahashi}; \cite{Kokubun}) data,
but the XIS data are used only in this paper because we perform a spatially resolved spectral analysis.
The XIS is an X-ray CCD camera that consists of one back-illuminated (BI) sensor (XIS1)
and three front-illuminated (FI) sensors (XIS0, XIS2, and XIS3). 
Each CCD covers a field of view of \timeform{18'} $\times$ \timeform{18'}, and
has a spatial resolution of about \timeform{2'} in half power diameter (HPD).
The BI sensor has a higher quantum efficiency in the soft energy band,
whereas the FI sensors have lower instrumental non X-ray background (NXB) in the hard energy band.

In this observation, the XIS2 data are not available because of the XIS2 anomaly
\footnote{http://www.astro.isas.ac.jp/suzaku/doc/suzakumemo/suzakumemo-2007-08.pdf}.
The XIS was operated in the normal clocking and 3 $\times$ 3 or 5 $\times$ 5 editing modes, and
the ``spaced-row charge injection'' (SCI: Uchiyama et al. 2007, 2009) option was applied.
Energy resolutions (full width at half maximum; FWHM) of FI and BI data in this observation
are 5.9 keV of $\sim$ 150 eV and $\sim$ 170 eV, respectively.

%%%%%% 2.2 Data reduction %%%%%%%%
\subsection{Data reduction}
\begin{figure}[h]
  \begin{center}
    \FigureFile(160mm,80mm){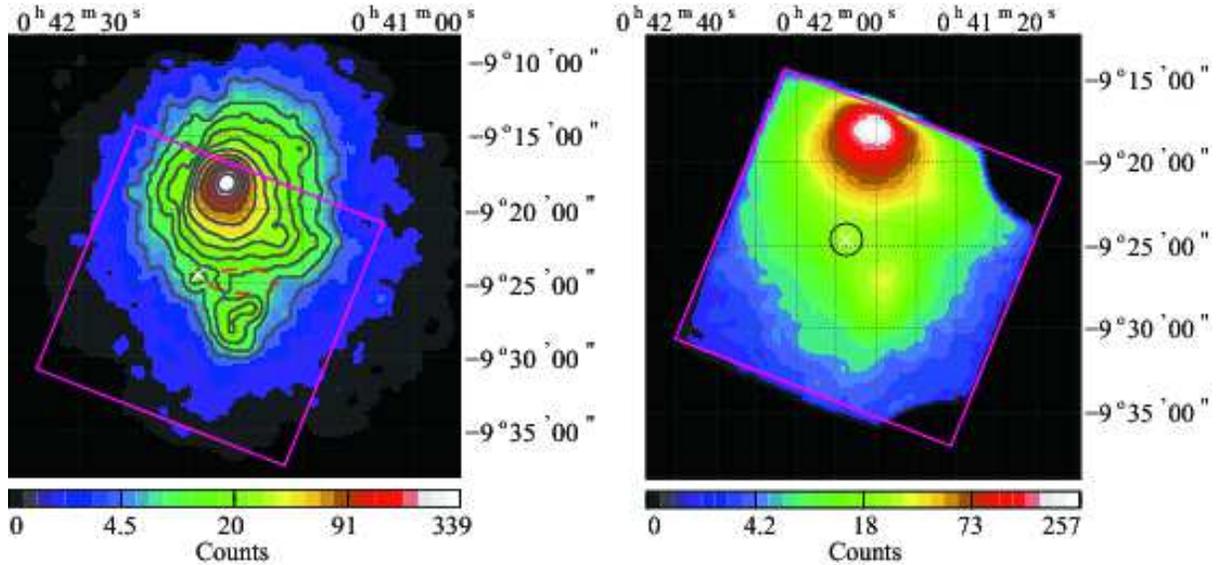}
  \end{center}
  \caption{
X-ray images of A 85.
Left: 
Image obtained with XMM-Newton/MOS1 in the 0.5--10 keV band
with background subtraction, but without any vignetting correction.
Smoothing over \timeform{30''} has been applied. 
Contour levels are logarithmically spaced. 
The magenta square represents the field of view of the Suzaku/XIS.
The orange dashed ellipse indicates the ``impact region'' defined in D05.
The white cross indicates the position of the X-ray point source ``DLF2005-17''.
Rright:
The Suzaku/XIS0 image in the 0.5--10 keV energy band.
Smoothing over \timeform{17''} has been applied. 
The white cross is the same as that for the left panel.
The black circle indicates the size of the Suzaku point spread function
(half power diameter of \timeform{2'}).
No background has been subtracted and no vignetting correction has been performed.
The calibration source regions are excluded.
}
\label{fig:xmm-img}
\end{figure}

We used reprocessed, unfiltered XIS event files (version 2.0) using HEASoft v6.5.1 (Suzaku software v10.0) and
the calibration database (CALDB) released on 2008 September 5.
The energy correction was performed by using ``xispi'', and
events were selected by using ``GRADE$=$0:0 2:4 6:6'' and ``STATUS$=$0:65535''.
Good time intervals (GTI) were determined by the standard criteria
referred to ``Updated Gain Calibration for SCI-on Data Nov 15, 2007''
\footnote{http://heasarc.gsfc.nasa.gov/docs/suzaku/analysis/sci\_gain\_update.html}.
To remove hot/flickering pixels, we applied ``cleansis'' to the event file.
The X-ray image generated from the cleaned event file is shown in the right panel of figure \ref{fig:xmm-img}.

For spectral analysis, redistribution matrix files (RMFs) and ancillary response files (ARFs)
were made by using ``xisrmfgen'' and ``xissimarfgen'' \citep{Ishisaki}.
The XMM MOS1 image of A 85 was used as the source distribution,
after subtracting the background, but not correcting the vignetting effect.
Using a different source image, we confirm that the vignetting correction
is insignificant in spectral analysis compared to the statistical errors on the results.
The background components, the NXB and the extragalactic cosmic X-ray background (CXB),
were estimated by the following procedure.
The NXB images and spectra were made by using ``xisnxbgen'' \citep{Tawa}.
For the CXB spectra, we referred to a method described in \citet{Sato}.
The ARF of a one-degree-radius uniform sky and the RMF were used to construct 
a CXB spectrum with the ``fakeit'' command of XSPEC (version 12.4.0ad).
The spectrum model of the CXB was a power-law model with photon index $\Gamma = 1.4$ and
$S_{\rm X}$[2--10 keV]$= 5.97 \times 10^{-8} \rm \ erg \ cm^{-2} \ s^{-1} \ sr^{-1}$,
where $S_{\rm X}$[2--10 keV] is the X-ray surface brightness in the 2--10 keV energy band
\citep{Kushino}, and $N_{\rm H} = 2.8 \times 10^{20} \rm \ cm^{-2}$ was adopted as
the hydrogen absorption column density \citep{Hartmann}.
Using the same requirement, the CXB events were created by ``xissim'' command to make the CXB images.

A bright point source was detected at the position of (RA, Dec) $=$ (\timeform{00h41m57s.5}, \timeform{-09D24'40''}) in the XMM image.
We call this source DFL2005-17, since it is named ``source 17'' in table 1 of D05.
The contribution of this point source in spectral analysis is insignificant because
the flux in the 2--10 keV energy band obtained from XMM-Newton data is less than 10\%
of the flux in Suzaku HPD, indicated by the black circle in the right panel of figure \ref{fig:xmm-img}.
Therefore, we neglect its contribution for spectral analysis hereafter.
Based on the ROSAT all-sky survey image (Snowden et al. 1995, 1997),
the surface brightness of the Galactic soft X-ray diffuse background around A 85 is relatively low.
Therefore, we apply a simple model for estimating this component.

%%%%% Figure 1 %%%%%%

%%%%%% 3. Analysis and Result %%%%%%%%
\section{Analysis and Results}
%%%%%% 3.1 Hardness ratio map %%%%%%%%
\subsection{Hardness ratio map}
\begin{figure}[h]
  \begin{center}
    \FigureFile(160mm,80mm){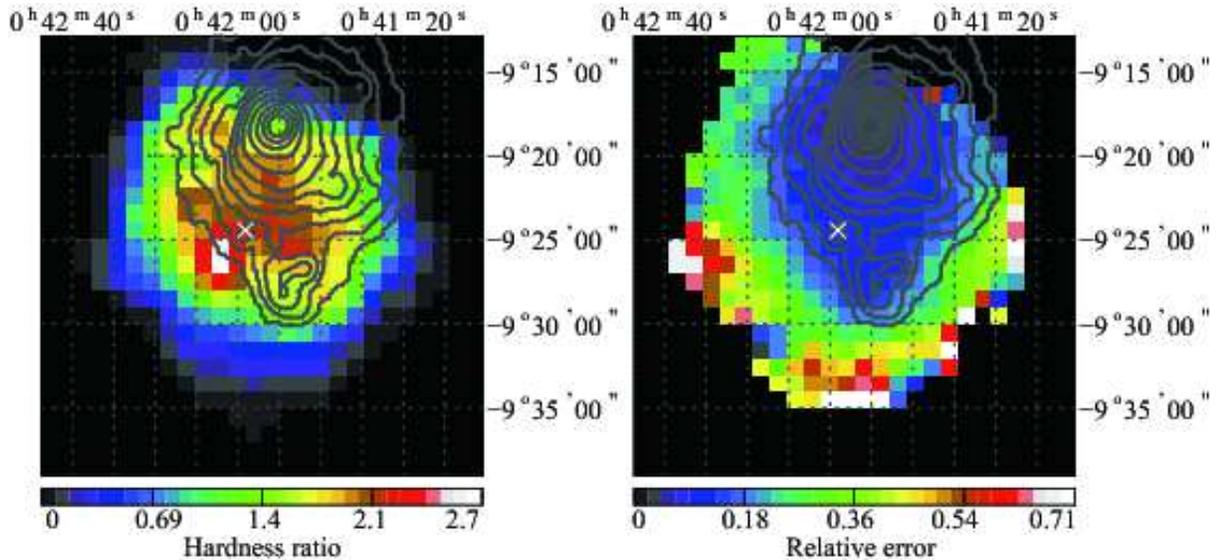}
  \end{center}
  \caption{
Hardness ratio (3--10~keV/0.5--2~keV; left) and relative error (right) maps of the FI.
The pixel width is \timeform{1'} and smoothing over \timeform{1'} has been applied.
The contours and the white cross are the same as those in figure \ref{fig:xmm-img}.
The background is subtracted without any vignetting correction.
%spell ok
}
\label{fig:hard}
\end{figure}

In this subsection, we generate an X-ray hardness ratio map and examine the overall temperature structure of the cluster.
The two FI CCD data (XIS0 and XIS3) are added together to increase the signal-to-noise ratio.
The BI data was not used in this analysis because of their higher backgrounds at high energy above 5 keV.
After subtracting the background components (NXB and the power-low CXB),
we made hard (3--10~keV) and soft (0.5--2~keV) X-ray images.
Figure \ref{fig:hard} shows the obtained hardness ratio and its relative error maps.

Between the main cluster and the subcluster, there is a hard region that was reported previously (D05).
Moreover, we found a much harder spot on the east side of the subcluster.
Note that this is $\sim$ \timeform{2'} away from the point source, DFL2005-17.
Hereafter, we refer to this spot as the hot spot.
Although a part of this hot spot may have been seen in the XMM-Newton results, 
main parts of the spot are not included in the analysis region of D05 (see figure 3 in D05).
The relative error map indicates that the detection of the hot spot has statistical significance
(with a relative error $<$ 15\%).
Therefore, the Suzaku observation detected this hot spot robustly for the first time.
%

%%%%% Figure 2 %%%%%%
%%%%%%%%%%%%%%%%%%%%

%%%%%% 3.2 Spectral analysis %%%%%%%%
\subsection{Spectral analysis}
%%%%%% 3.2.1 Temperature of each region %%%%%%%%
\subsubsection{Temperature of each region}
\begin{figure}[h]
  \begin{center}
    \FigureFile(160mm,80mm){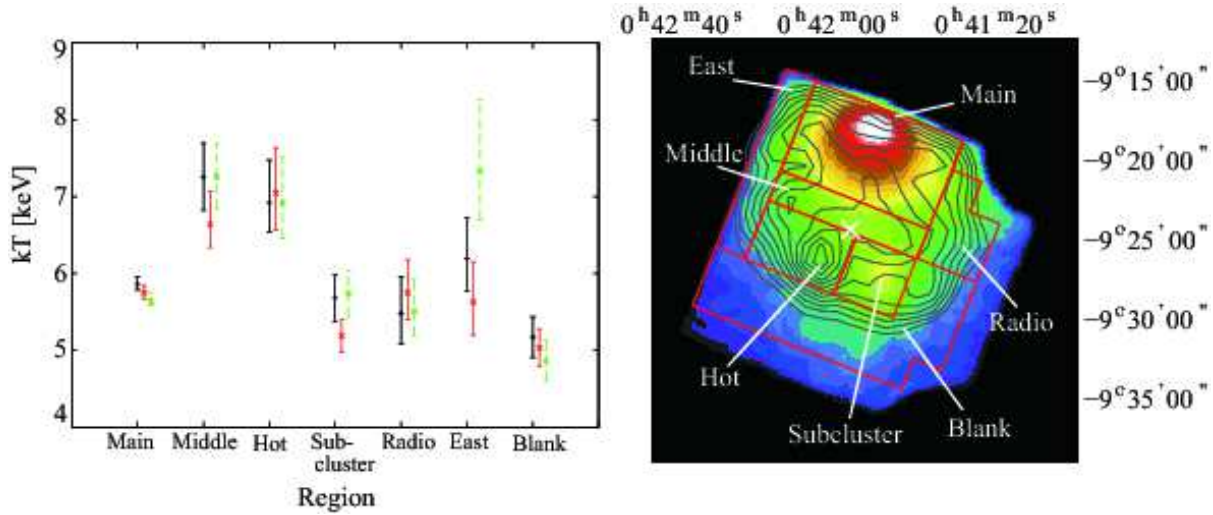}
  \end{center}
  \caption{
Left:
Temperature profiles of the XIS0 (black), XIS1 (red), and XIS3 (green) detectors.
The horizontal axis represents regions extracted spectra. 
Right:
The spectral extraction regions are indicated on the XIS0 image.
The contours of the hardness ratio map (the left panel of figure \ref{fig:hard}) are also overlaid.
The ``Radio'' region includes some radio sources (\cite{Bagchi}; \cite{Lima2001}).
The white cross is the same as that in figure \ref{fig:xmm-img}.
}
\label{fig:region-graph}
\end{figure}

We divided the field of view into seven regions based on the hardness ratio map,
as shown in the right panel of figure \ref{fig:region-graph}.
We extracted the X-ray spectrum from these seven regions separately.

The energy ranges used for the spectral analysis were 0.5--10 keV for FI and 0.5--7 keV for BI.
The energy range of the anomalous response around the Si K-edge (1.825--1.840 keV) has been excluded in both cases.
To describe the cluster emission, we used a collisional ionization equilibrium model (APEC; \cite{Smith})
modified by the Galactic photoelectric absorption (wabs in the XSPEC), using XSPEC.
The Galactic hydrogen column density was fixed to the reported value, 2.80 $\times 10^{20} \ \rm cm^{-2}$ \citep{Hartmann}.
To model the Galactic soft X-ray background, another APEC model was added to the model.
By fitting the spectrum in ``Blank'' region, temperature, $kT$, is determined as 0.18~keV.
This value was used for other regions.
In contrast, normalizations of other regions are left free to compensate the variation of the background flux.

\begin{figure}[h]
  \begin{center}
    \FigureFile(150mm,70mm){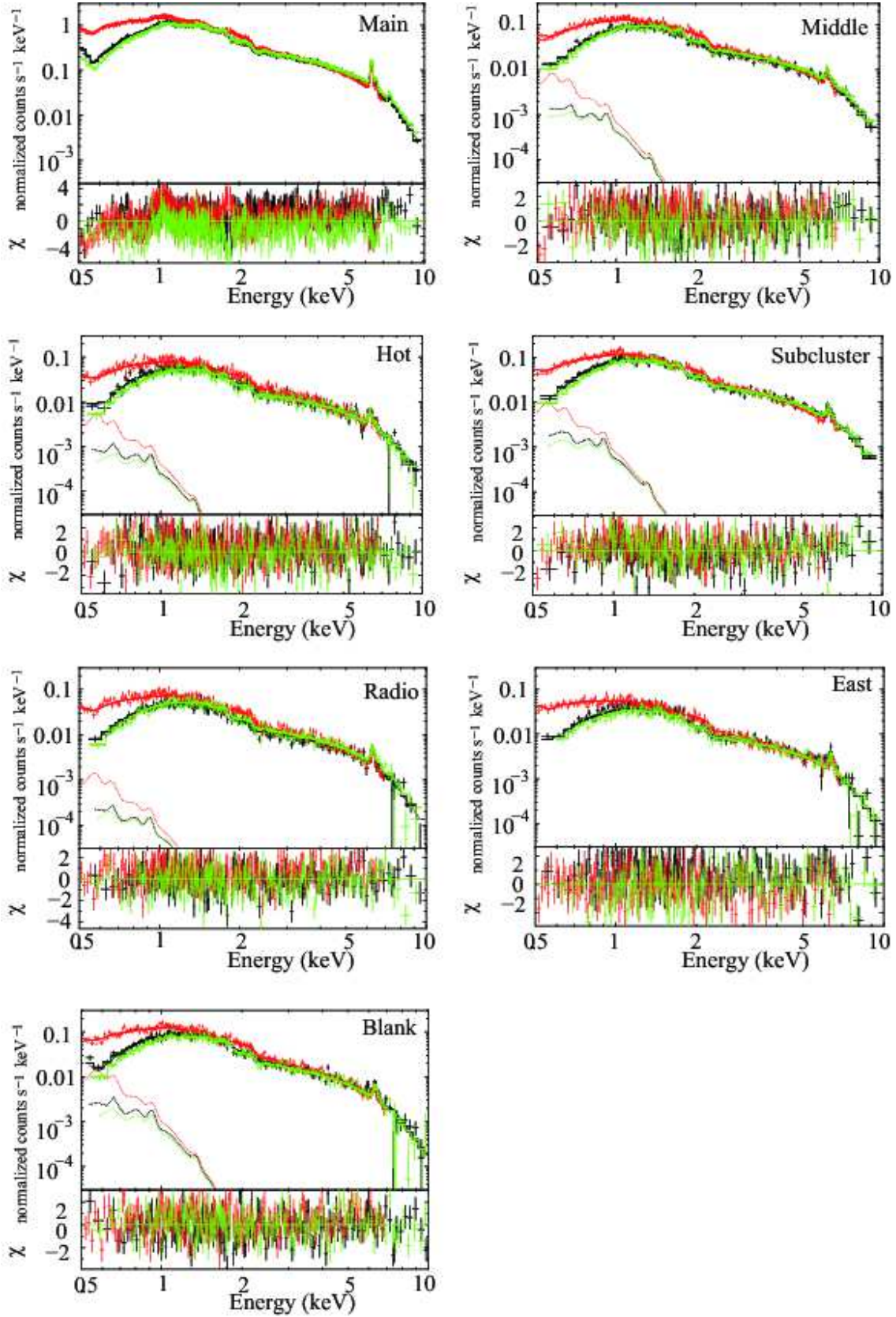}
  \end{center}
  \caption{
Results of spectral fitting.
Each panel shows the background (the power-low CXB $+$ NXB) subtracted spectra
of the region indicated in the right panel of figure \ref{fig:region-graph}.
Black, red, and green crosses indicate XIS0, XIS1, and XIS3 data, respectively.
In the upper part of panel,
the solid lines indicate the best-fit model for ``APEC $\times$ wabs $+$ APEC$_{G}$'', and
the lower dotted lines show the component ``APEC$_{G}$'' of the model, as the Galactic soft X-ray component.
Note that in ``Main''- and ``East''- region fitting,
the components do not appeared in each plot because of the low flux.
In the lower part of each panel, the fit residuals are shown in units of $\sigma$.
}
\label{fig:reg-spec}  
\end{figure}

Figure \ref{fig:reg-spec} shows the fitting result of each region
and table \ref{tab:para} summarizes the obtained parameters.
The fitting from the BI gave larger statistical errors than
those from FI, in general.
The spectral fitting of ``Main'' region indicates an excess at an energy of around 1.1 keV,
as shown in upper-left panel in figure \ref{fig:reg-spec}.
Adding another thermal component, we could improved the fit.
This is possibly because of the cool component commonly seen in the center of galaxy clusters (e.g. \cite{Fabian}).

The left panel of figure \ref{fig:region-graph} shows
the temperature profiles obtained by using the three detectors.
The temperature of ``East'' region from the XIS3 is inconsistently
higher than that obtained from other detectors.
This is possibly caused by a calibration error at the edge of CCDs.
We omit any discussion of ``East'' region hereafter.
The ``Middle'' and ``Hot'' regions are significantly hotter ($kT \sim 7$ keV)
than the ``Main'' and ``Subcluster'' regions ($kT \sim 6$ keV). 
This is consistent with the hardness ratio map (figure \ref{fig:hard}).
A more detailed temperature distribution is given in the next subsection.

%%%%% Figure 3, 4 %%%%%%

%%%%% Table 1 %%%%%%
\begin{table}[h]
\begin{center}
\caption{
{\normalsize 
Results of spectral fitting for the seven regions,
using the added FI (XIS0 $+$ XIS3) spectra.}
}
\label{tab:para}
\scalebox{0.9}[1.0]{
\begin{tabular}{lccccccc}
\hline\hline
Region & 
Main   & 
Middle & 
Hot    & 
Subcluster &
Radio  & 
East   & 
Blank \\ \hline
$kT$(keV) & 
$5.73^{+0.06}_{-0.05}$ & % Main, ok
$7.23^{+0.32}_{-0.31}$ & % Middle, ok
$6.87^{+0.41}_{-0.28}$ & % Hot, ok
$5.76^{+0.22}_{-0.21}$ & % Sub, ok
$5.58^{+0.30}_{-0.28}$ & % Radio, ok
$6.75^{+0.50}_{-0.32}$ & % East, ok
$5.07^{+0.18}_{-0.18}$   % Blank, ok
\\ \hline
Redshift & 
$0.0562^{+0.0005}_{-0.0005}$ & %Main, ok
$0.0540^{+0.0026}_{-0.0025}$ & %Middle, ok
$0.0562^{+0.0031}_{-0.0041}$ & %Hot, ok
$0.0581^{+0.0036}_{-0.0026}$ & %Sub, ok
$0.0565^{+0.0048}_{-0.0046}$ & %Radio,ok 
$0.0569^{+0.0057}_{-0.0027}$ & %East, ok
$0.0603^{+0.0060}_{-0.0051}$   %Blank, ok
\\ \hline
$Z$(solar) & 
$0.46^{+0.02}_{-0.02}$ & %Main, ok
$0.27^{+0.05}_{-0.05}$ & %Middle, ok 
$0.26^{+0.06}_{-0.06}$ & %Hot, ok
$0.21^{+0.04}_{-0.04}$ & %Sub,ok
$0.26^{+0.06}_{-0.06}$ & %Radio, ok
$0.36^{+0.10}_{-0.09}$ & %East, ok
$0.17^{+0.04}_{-0.04}$   %Blank, ok
\\ \hline
Norm$^{*, \dagger}$ & 
$38.0^{+0.2}_{-0.1}$ &   %Main, 
$2.25^{+0.03}_{-0.03}$ & %Middle, ok
$1.30^{+0.03}_{-0.03}$ & %Hot, ok
$2.44^{+0.04}_{-0.04}$ & %Sub, ok
$1.52^{+0.04}_{-0.03}$ & %Radio, ok
$1.17^{+0.02}_{-0.03}$ & %East, ok
$2.38^{+0.05}_{-0.05}$   %Blank, ok
\\ \hline
$\chi^2$/d.o.f. & 
1368/938 & %Main, ok
356/303 & %Middle, ok
416/362 & %Hot, ok
309/276 & %Sub, ok
339/311 & %Radio, ok
234/249 & %East, ok
305/303   %Blank, ok
\\ 
\hline
\multicolumn{4}{@{}l@{}}
{\hbox to 0pt{\parbox{160mm}
{\footnotesize
\par\noindent
\footnotemark[$*$] The APEC model normalization is $10^{-17} \int n_e n_H dV / (4 \pi D_{A} (1 + z)^{2})$.
\par\noindent
\footnotemark[$\dagger$] Normalizations are rescaled multiplying by $\rm SOURCE\_RATIO\_REG$ in
FITS header of the calculated ARFs using ``xissimarfgen''.
%\par\noindent
     }\hss}}
   \end{tabular}
   }
 \end{center}
\end{table}

%%%%%% 3.2.2 Detailed temperature map  %%%%%%%%
\subsubsection{Detailed temperature map}
In order to investigate the detailed temperature distribution of the cluster,
we made a fine temperature map focusing on the hot component in particular.
After excluding the ``Blank'' region with low S/N,
we divided the whole region into \timeform{2'} $\times$ \timeform{2'} boxes.
We perform spectral fitting in the similar way as discribed the previous subsection,
but only used the FI spectra.

The half power diameter of the telescope point-spread function, \timeform{2'},
is comparable to the spatial bin size in our analysis.
We estimated that about 50\% of the event in a box originated from surrounding regions (see also \cite{Ota}).
Therefore, our obtained temperature map was substantively diluted, and
the actual temperature variation could be larger than the present result.
The reduced $\chi^2$ of spectral fittings in all boxes are in the range of 0.86--1.20,
except for CCD edge regions (boxes 00--50).

\begin{figure}[h]
  \begin{center}
    \FigureFile(160mm,80mm){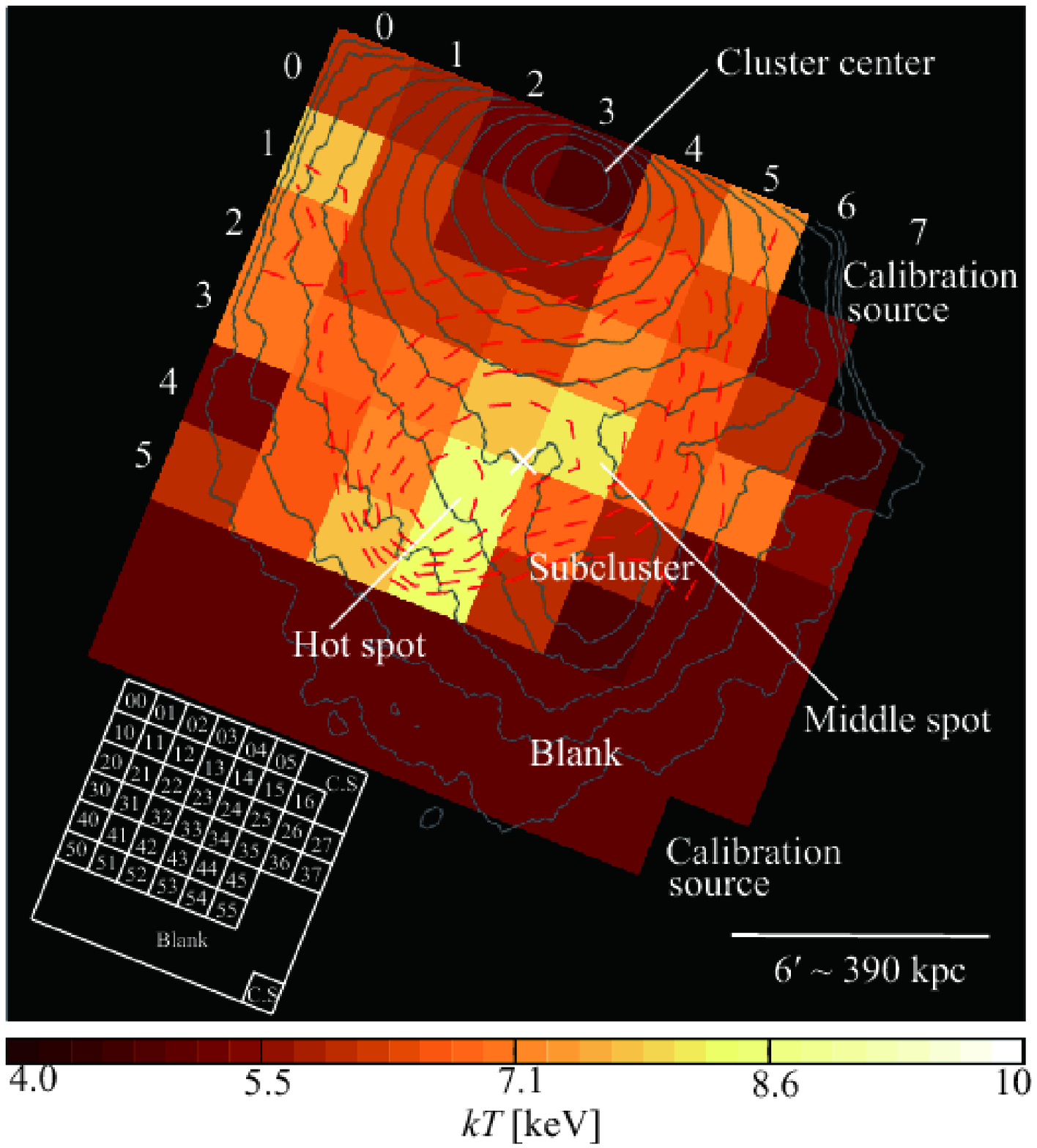}
  \end{center}
  \caption{
Temperature distribution map.
The contours of X-ray surface brightness from the Suzaku data are shown by black lines on a logarithmic scale.
The contours of hardness ratio are indicated by red dashed lines on a liner scale.
Calibration sources located at northwest and southwest have been excluded.
The white cross is the same as that in figure \ref{fig:xmm-img}.
The box address map is given in the bottom-left, where ``CS'' represents the positions of the calibration source regions. 
}
\label{fig:temp-map} 
\end{figure}

The obtained temperatures are shown in figure \ref{fig:temp-map}.
The cluster center (box 03) and the subcluster (boxes 44, 45, 54 and 55) are cool
($kT \sim 4.8$ keV and $\sim$ 4.9--6.8 keV, respectively)
relative to a high-temperature spot between the main part of the cluster and subcluster,
(box 34, $kT \sim 8.2$ keV, hereafter we refer to this spot as the middle spot)
and the hot spot (boxes 43, $kT \sim 8.5$ keV).
The low temperatures at the cluster center and the outer envelope (``Blank'' region)
may reflect the general temperature profile of galaxy clusters (e.g. \cite{Allen}).
This temperature distribution is consistent with the hardness ratio map obtained in \S 3.1.

\begin{figure}[h]
  \begin{center}
    \FigureFile(160mm,80mm){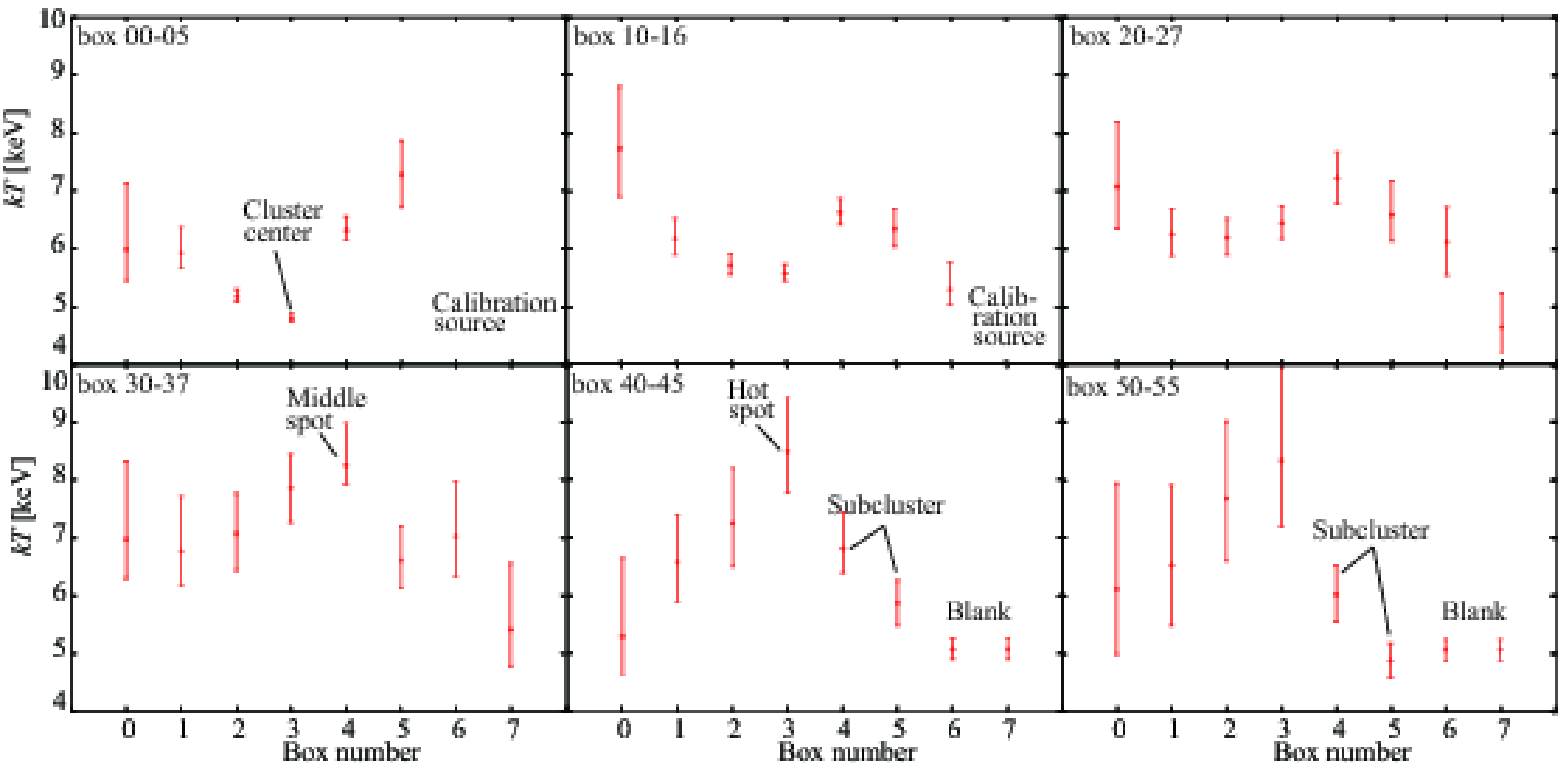}
  \end{center}
  \caption{
One-dimensional temperature profiles along the east-west direction.
The box addresses are shown in figure \ref{fig:temp-map}.
}
\label{fig:box-graph} 
\end{figure}

The same results are shown in figure \ref{fig:box-graph}
in one-dimensional plots along the detectors' axes.
The temperature profile of ``boxes 40--45'' in the figure shows
a significant temperature difference of $\sim$ 1.7 keV 
between boxes 43 (hot spot) and 44 (a part of the subcluster).
Other temperature difference can be seen between boxes 03 and 04, 
between boxes 13 and 14, and between boxes 53 and 54.
Some of the temperature variations are already shown in figure 3 of D05,
except for the difference between boxes 53 and 54.
This analysis have revealed the temperature structure including the hot spot very robustly,
and in wider spatial area than in previous measurements.

%%%%% Figure 5, 6 %%%%%%

%%%%%% 3.2.3 Maps of other quantities %%%%%%%%
\subsubsection{Maps of other quantities}
\begin{figure}[h]
  \begin{center}
    \FigureFile(160mm,80mm){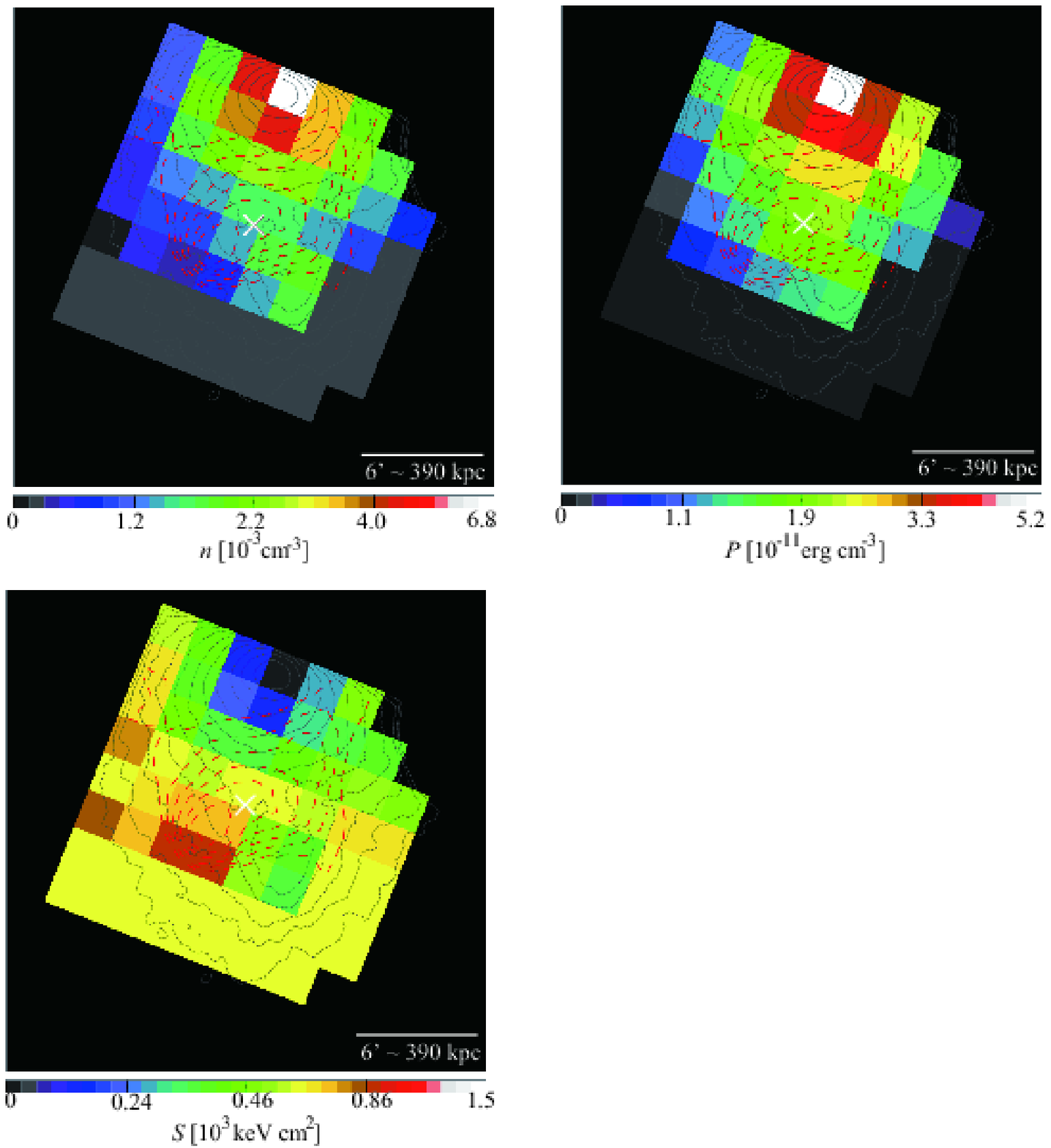}
  \end{center}
  \caption{
Particle number density, pressure and entropy maps are shown in top left,
top right and bottom left, respectively.
The contours are the same as those in figure \ref{fig:temp-map}, and
the white cross is the same as that in figure \ref{fig:xmm-img}.
}
\label{fig:box-maps}
\end{figure}

In search for further evidence of the merging of subsystems and features in A 85,
we produced projected maps of particle number density, pressure, and 
entropy index, as shown in figure \ref{fig:box-maps}.
The electron number density is calculated from the normalization of the APEC,
$K = 10^{-14} \int n_{\rm H} n_e dV / (4 \pi D_A^2(1+z)^2) ({\rm cm^{-5}}$),
where $n_e$ and $n_{\rm H}$ are, respectively, the electron and hydrogen densities in units of $\rm cm^{-3}$
and $D_A$ is the angular diameter distance in units of cm.
The solid angle $\Delta \Omega$ of each box is 4 arcmin$^2$.
Then, projected $n_e$ is obtained as \citep{Henry}
\begin{eqnarray}
n_e = 7.28 \times 10^{-7} (1 + z) \left( \frac{D_A}{1 {\rm Mpc}} \right) 
\left( \frac{K}{10^{-3} \rm cm^{-5}} \right)^{1/2} \left(\frac{V}{1 \rm Mpc^3} \right)^{-1/2} \ \rm cm^{-3}.
\end{eqnarray}
Here, we assumed that the volume is $V \sim (2/3) \ D_A^2 \ \Delta \Omega \left( R_{\rm vir}^2 - R^2  \right)^{1/2}$,
where $R$ is the projection radius from the center of the cluster and $R_{\rm vir}$ is the virial radius, $\sim 2$ Mpc \citep{Reiprich}.
By using H:He $=$ 9:1 in terms of numbers of atoms as the abundance ratio of the ICM,
the total particle (electron plus ion) number density is obtained to be $n = (13/11 \mu ) n_e$ \citep{White},
where $\mu \sim 0.62$ is the mean molecular weight.
The pressure is given by $P = nkT$, and we use the entropy index $S \equiv kTn^{-2/3}$
instead of the thermodynamic definition $(3k/2) {\rm ln} [T \rho_{\rm g}^{-2/3}]$,
where $\rho_{\rm g}$ is the gas density.

The number density map naturally resembles the X-ray surface brightness distribution.
The high-temperature region (boxes 33, 34, 43 and 53) has a relatively high pressure.
The east side of the subcluster (boxes 42, 43, 52 and 53) has a higher entropy than the surrounding region.
The entropy in this region might be increased by merging.
These maps of particle number density, pressure and entropy are used in the discussion below.
%

%%%%% Figure 7 %%%%%%

%%%%%% 3.3 Complementary analysis using optical data %%%%%%%%
\subsection{Complementary analysis using optical data}
Although we expected that a significant in difference redshift between
the main part of A 85 and the subcluster would be detected in the X-ray observation,
it was not obtained (see table \ref{tab:para}).
Then, we consider that the difference may be detected by using the radial velocities
of the member galaxies obtained from the optical spectroscopic observations
whose accuracy of the redshifts are higher than those derived from X-ray data. 
In addition, we may confirm other features of the member galaxies
using such information as optical colors and spatial distribution.

The catalog of \citet{Durret1998a} is used for our analysis.
This catalog includes the information on the positions and the radial velocities 
of member galaxies obtained from their optical spectroscopic observation.
The catalog covers about 1 square degree of cluster center, and
the typical error on radial velocities in this catalog is smaller than 100 km s$^{-1}$.
Though the catalog does not cover homogeneously the whole region,
it is enough for identifying physical concentration of galaxies around main and subcluster regions,
which is essential for deriving mean radial velocities of both components.
Since the catalog includes not only early-type, but also late-type galaxies,
we can perform the color analysis for the member galaxies
by adding the color information to the galaxies.
For the purpose, we matched the catalog with the photometric catalog of
the Sloan Digital Sky Survey (SDSS; \cite{York}) DR-7 \citep{Abazajian}
by identifying the SDSS galaxies within the position of $\pm 2$ arcsec.
In this subsection, we use that data set hereafter.

\begin{figure}[h]
  \begin{center}
    \FigureFile(160mm,80mm){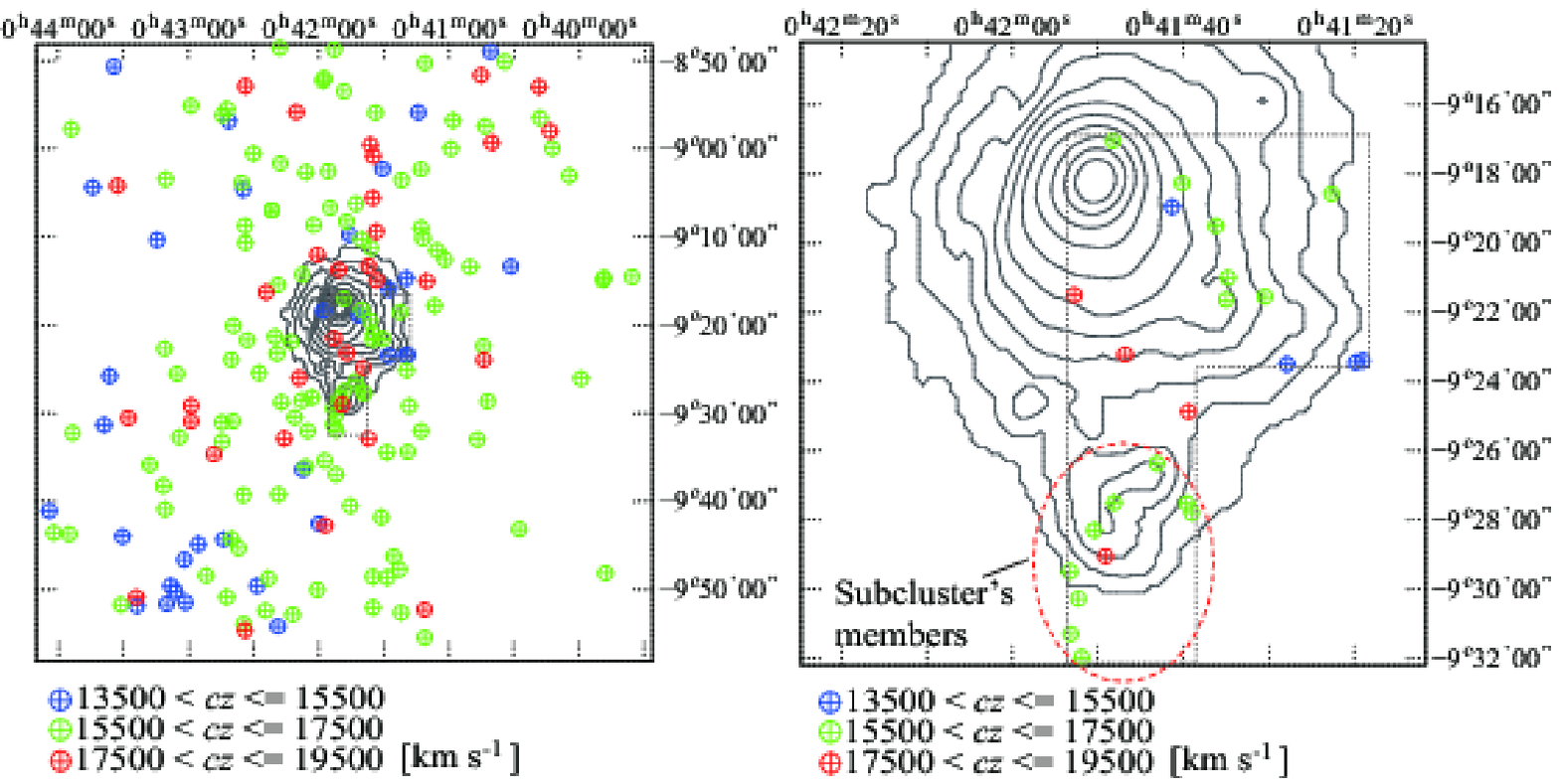}
  \end{center}
  \caption{
Left: Spatial distribution of member galaxies with radial velocity.
The contour levels are the same as those in the left panel of figure \ref{fig:xmm-img}.
The part enclosed by dashed lines is enlarged in the right panel.
Right:
Close-up of the left panel. The points and the contours are the same as those in the left panel.
It is conjectured that galaxies enclosed by red dashed lines are the member galaxies of the subcluster, listed in table \ref{tab:sub-member}.
}
\label{fig:cz-map} 
\end{figure}

The spatial distribution of the member galaxies is shown in the left
panel of figure \ref{fig:cz-map}.
We considered that the member galaxies of A 85 should have radial velocities $cz$ within 13500--19500 km s$^{-1}$
because the systemic velocity of A 85 is 16507 $\pm$ 102 km s$^{-1}$ \citep{Oegerle}.
To confirm the radial velocities of individual regions visually,
we separate the member galaxies into three radial-velocity groups.
Blue, green and red points indicates the range of the radial velocity of
13500--15500, 15500--17500 and 17500--19500 km s$^{-1}$, respectively.
Because of its compact concentration and positional coincidence,
we here considered that the 10 galaxies shown in the right panel of figure \ref{fig:cz-map}
surrounded by dotted line are the member galaxies of subcluster.
In the region between the center and the subcluster,
there are some galaxies with relatively large radial velocities, shown by red points. 
It might be indicative that there was another collision by small group of galaxies previously in this region.
Southeast of the subcluster, there is a clustering of galaxies with small radial velocities,
which is consistent with the report that an X-ray filament expanding in this direction \citep{Durret2003}.
The member galaxies of the subcluster (see the right panel of figure \ref{fig:cz-map}) listed in table \ref{tab:sub-member}
have almost the same radial velocities ,and their mean radial velocity is $17028 \pm 216$ km s$^{-1}$.
This value is lager than the systemic velocity of A 85 $\sim 520 \pm 240$ km s$^{-1}$,
and it is consistent with the fact that a significant difference radial velocity could not be detected by this X-ray observation. 

\begin{figure}[h]
  \begin{center}
    \FigureFile(160mm,80mm){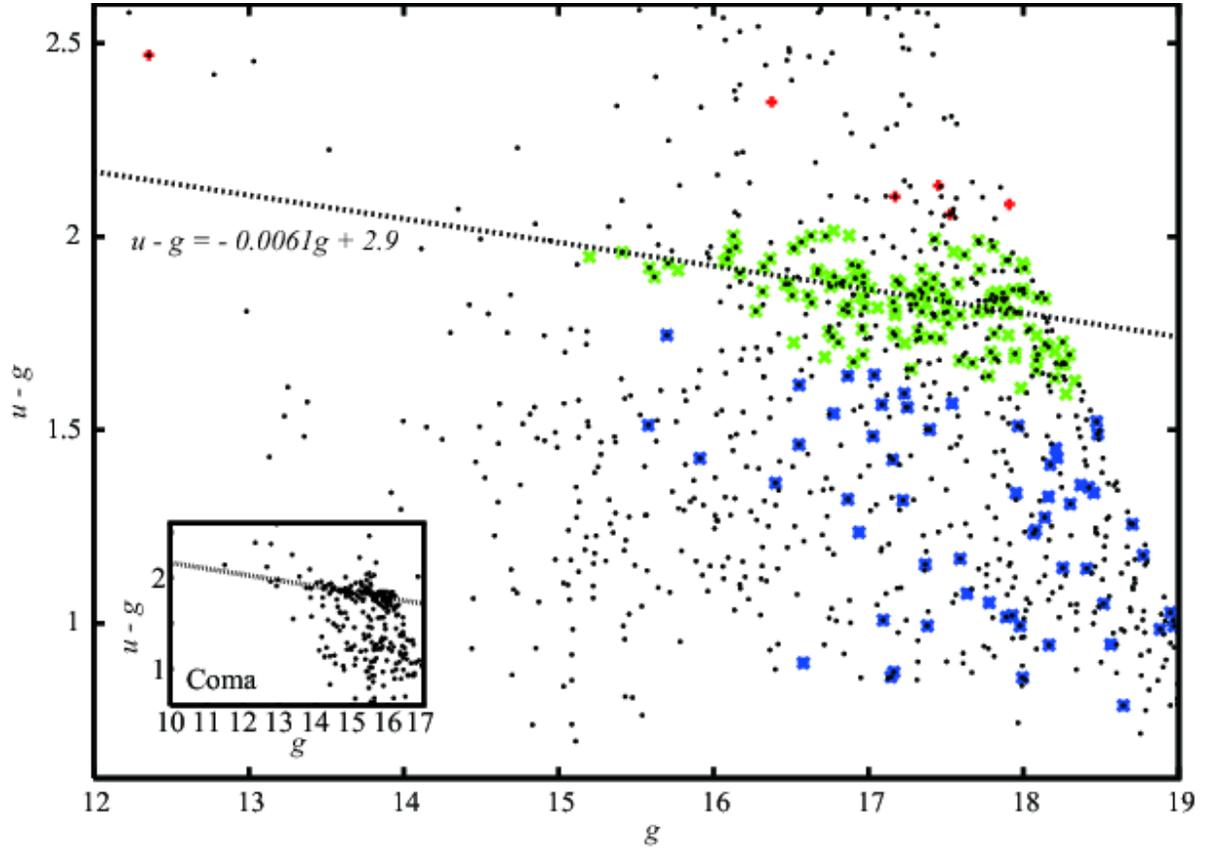}
  \end{center}
  \caption{
Color-magnitude diagram, $u-g$ versus $g$, of A 85 field.
The dotted line represents the relation $u-g = -0.061 g + 2.9$.
Green crosses indicate galaxies distributed along the relation within the $u-g$ color difference of $\pm$ 0.2 mag.
Red and blue crosses represent galaxies that are redder and bluer than those shown by green points, respectively. 
Black points indicate the photometric data of SDSS DR7 within the field covered by \citet{Durret1998a}.
A similar diagram for Coma cluster is shown in the small panel of the left bottom.
In the small panel, points indicate photometric data of SDSS DR7 in Coma cluster,
and the dotted line represents the relation $u-g = -0.064 g + 2.8$.
}
\label{fig:color-map}  
\end{figure}

The diagram, $u-g$ versus $g$, of the member galaxies is shown in figure \ref{fig:color-map}.
As discussed in \citet{Bower}, there must be ``red-sequence'' made by early-type member galaxies in rich clusters of galaxies.  
From the distribution of the member galaxies, we can derive the relation, $u-g = -0.061 g + 2.9$ (dotted lines),
and it is very consistent with that of Coma cluster (see left bottom of figure \ref{fig:color-map}). 
The facts shown in this figure are as follows:
(1) From the photometrical data of SDSS-DR7 only, it is difficult to find the ``red-sequence''.
As there are some reports that some components are overlapping along the line-of-sight in the field of A 85
(e.g. \cite{Bravo-Alfaro}), the ``red-sequence'' might not clearly be seen. 
(2) Many of the member galaxies gather near the relation,
while there exist a certain number of galaxies below them (bluer than them).
This implies that A 85 is enriched with blue galaxies, which may be proceeding
more actively with star formation than typical rich clusters.
On the other hand, we could not get any remarkable features in the spatial
distribution of the member galaxies with color information.

%%%%% Figure 8, 9 %%%%%%

%%%%% Table 2 %%%%%
\begin{table}
  \caption{Coordinates, magnitudes of the $u$ and $g$ bands, $u-g$, and $cz$ with errors for the member galaxies of the subcluster.}
  \label{tab:sub-member}
  \begin{center}
    \begin{tabular}{ccccccc}
      \hline\hline
RA & Dec & $u$ & $g$ & $u-g$ & $cz$ \\
(J2000.0) & (J2000.0) &  &  &  &  (km s$^{-1}$) \\
\hline
00 41 39.0 & -09 27 48 & 19.99 & 18.30 & 1.69 & 17257$\pm$52 \\
00 41 39.6 & -09 27 31 & 18.58 & 17.16 & 1.42 & 16647$\pm$62 \\
00 41 43.1 & -09 26 22 & 17.37 & 15.41 & 1.96 & 16886$\pm$35 \\
00 41 48.1 & -09 27 30 & 19.49 & 17.79 & 1.70 & 17203$\pm$66 \\
00 41 49.1 & -09 29 03 & 19.78 & 17.80 & 1.98 & 18437$\pm$136 \\
00 41 50.4 & -09 28 18 & 19.93 & 18.09 & 1.84 & 17293$\pm$49 \\
00 41 51.9 & -09 31 58 & 19.45 & 17.58 & 1.87 & 16523$\pm$243 \\
00 41 52.3 & -09 30 16 & 17.51 & 15.62 & 1.89 & 17164$\pm$33 \\
00 41 53.2 & -09 31 17 & 19.56 & 17.81 & 1.75 & 17121$\pm$67 \\
00 41 53.3 & -09 29 29 & 18.01 & 17.14 & 0.87 & 15751$\pm$75 \\
      \hline
    \end{tabular}
  \end{center}
\end{table}

%%%%%% 4 Discussion %%%%%%%%
\section{Discussion}
%%%%%% 4.1 Impact direction %%%%%%%%
\subsection{Impact direction}
Previous studies suggest that the subcluster has moved from the southeast of the present position, based on the following findings:
(1) X-ray observations show that the cold front is located between the subcluster and the main part of A 85 (Chandra, \cite{Kempner}). 
(2) An X-ray filament extends from the subcluster to the southeast direction (XMM--Newton, \cite{Durret2003}).
(3) The ``impact region'' has a high temperature (XMM--Newton, \cite{Durret2005}).
(4) Some H$\alpha$ emitting galaxies are in the southeast of the subcluster \citep{Boue}.

\begin{figure}[h]
  \begin{center}
    \FigureFile(160mm,80mm){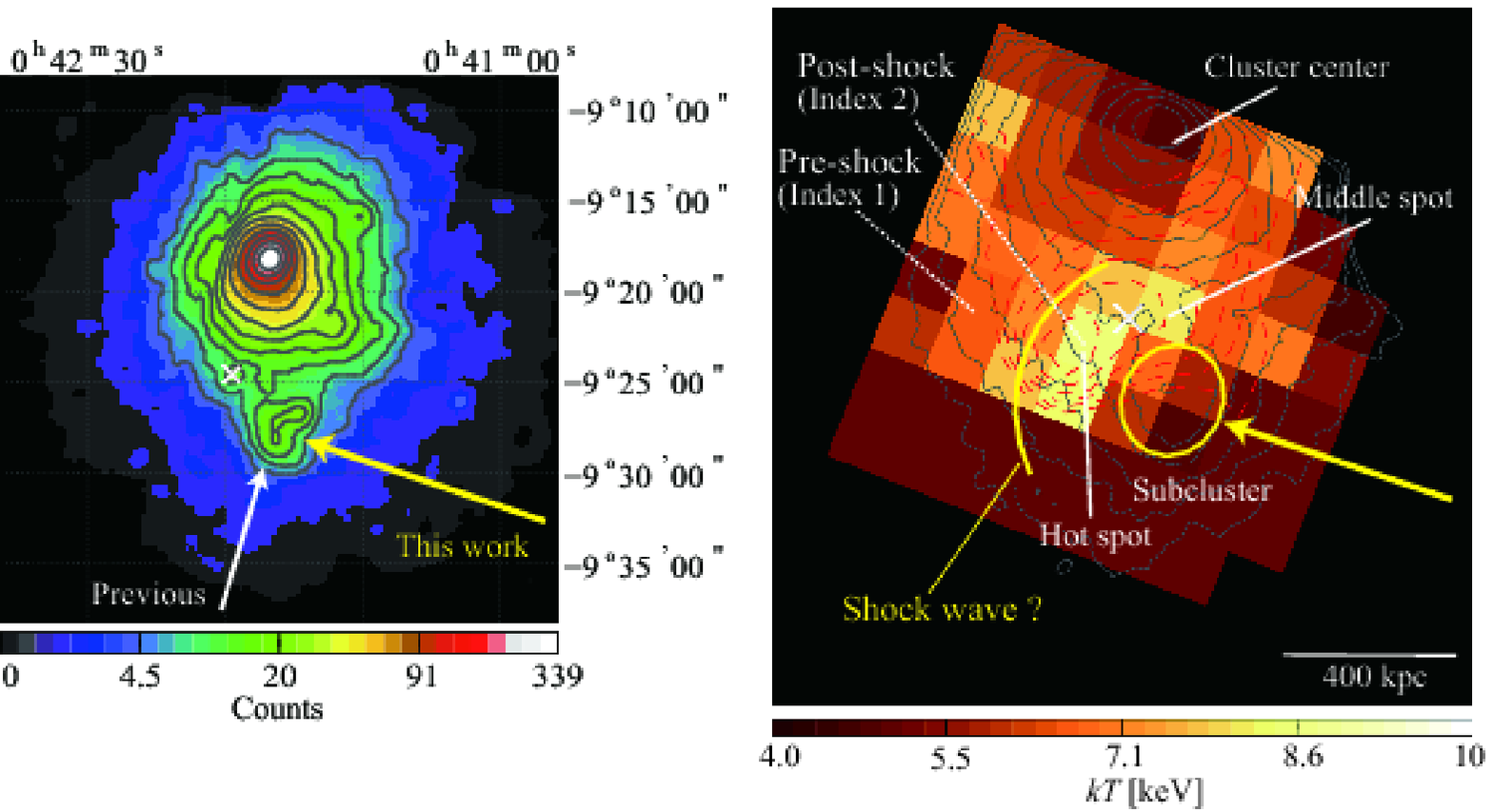}
  \end{center}
  \caption{
Left:
Impact directions of the subcluster proposed in previous studies (white arrow)
and in this study (yellow arrow) superimposed on the XMM-Newton image.
Right:
Bird's eye view of the shock wave and the motion of the subcluster.
The yellow arc indicates the expected location of the shock front and the ellipse indicates the subcluster.
The temperature map is the same as that in figure \ref{fig:temp-map}.
}
\label{fig:ponchi}
\end{figure}

However, we found a hot spot in the east side of the subcluster.
This feature can not be easily explained by merging of the subcluster from the southeast direction.
Instead, it is more plausible to consider that this hot spot is caused by merging from the southwest direction.
Therefore, we proposed and examined a new model in which the subcluster collides from the southwest
(see the left panel of figure \ref{fig:ponchi}).
Regarded from this new point of view, the ``impact region'' appearing in D05 is considered to be the northwest end of the shocked region.
Moreover, the listed facts wiht which the collision from the southeast direction
is observationary supported do not critically conflict with merging from the southwest direction.

%%%%% Figure 10 %%%%%%

%%%%%% 4.2 Motion of the subcluster %%%%%%%%
\subsection{Motion of the subcluster}
Here, we estimate the infall (merging) velocity of the subcluster using
Rankine--Hugoniot jump conditions (RH conditions)
in the case of merging with the subcluster from the southwest direction.
To obtain the information about the motion of the subcluster,
we consider the boxes 41 and 43 to be the typical parts of
preshock (index~1) and postshock regions (index~2), respectively
(see the right panel of figure \ref{fig:ponchi}).

The way we consider this problem is as follows:
(I) ``The shock-wave speed is equal to the infall velocity of the subcluster.''
Since the shock wave is formed in the open system,
we consider that the whole structure of the shock wave plus the subcluster keep a steady form,
i.e. the shock-wave speed is equal to the present velocity of the subcluster relative to A 85.
(II) ``The postshock region is consists of two temperature components.''
The results obtained in previous sections are physical quantities integrated along the line of sight.
In order to obtain deprojected quantities, we fit the two-temperature model ((APEC $+$ APEC) $\times$ wabs $+$ APEC$_G$)
to the spectrum in the postshock region (box 43).
We consider that the quantities of high-temperature ($kT_2 = 10.4^{+2.9}_{-1.9}$ keV) component are those of the postshock region.
The extent and the depth of the postshock region are estimated to be 4 arcmin$^2$ and 5--8 arcmin
when the shock wave passed through the region.
This gives the deprojected particle number density as $n_2 = 1.70 \pm 0.20 \times 10^{-3}$.
The estimations of the volume ot the postschok region are consistent with the emission integral of the region.
(III) ``The initial conditions of the postshock region are equal to those of the preshock region.''
Regions in the same radius would be similar physical state in relaxed galaxy clusters before the merging.
Since the postshock and the preshock regions are seem to be at almost equal distance ($\sim \timeform{8'}$) from the center of A 85,
we consider that initial conditions in these regions were nearly equal.

Using the obtained temperature and the particle number density of the postshock region,
we estimate the Mach number related with the merging, and the infall velocity of the subcluster.
We later note that the infall velocity is somewhat affected by a possible breaking of assumption (III)
as the difference in the initial conditions between the preshock and the postshock region before the merging.

By using RH conditions, the following relation is obtained \citep{Landau}:
\begin{eqnarray}
\frac{P_2}{P_1} = \frac{2 \gamma }{\gamma + 1} \mathcal{M}_1^2 -
\frac{\gamma - 1}{\gamma + 1},
\end{eqnarray}
where $\gamma$ = 5/3 is the adiabatic index for the fully ionized plasma and
$\mathcal{M}_1 = v_1 / c_{\rm s1}$ is the Mach number, and
where $v_1$ and $c_{\rm s1}$ are the infall velocity of the subcluster and
the sound speed in the preshock gas, respectively.
$P_1$ and $P_2$ are given by $P_1 = 1.09^{+0.14}_{-0.12} \times 10^{-11}$ erg cm$^{-3}$
($kT_1 = 6.57^{+0.82}_{-0.67}$ keV, $n_1 = 1.04 \pm 0.03 \times 10^{-3}$ cm$^{-3}$),
and $P_2 = 2.82^{+0.85}_{-0.61} \times 10^{-11}$ erg cm$^{-3}$
($kT_2 = 10.4^{+2.9}_{-1.9}$ keV, $n_2 = 1.70 \pm 0.20 \times 10^{-3}$ cm$^{-3}$).
We then obtain $P_2/P_1 = 2.59^{+0.85}_{-0.63}$, $c_{\rm s1} = 1300^{+80}_{-70}$ km s$^{-1}$;
$\mathcal{M}_1$ and $v_1$ are estimated as $1.5 \pm 0.2$ and $1950^{+290}_{-280}$ km s$^{-1}$.
By using the difference redshift between the subcluster and A 85, $520 \pm 240$ km s$^{-1}$ obtained in \S 3.3,
the angle between the line of sight and the direction of motion of the subcluster is found to be
$\cos^{-1} (520 \pm 240/1950^{+290}_{-280}) \simeq 75^{+7}_{-8}$ degrees.
Here, we estimate the effects of a possible breaking of the assumption (III).
From $v_1 = \mathcal{M}_1 c_{\rm s1}$, we found that $v_1$ mainly depends on $(P_2/n_1)^{1/2}$.
Therefore, for example, $v_1$ decreases (increases) by 5\% if $n_1$ increases (degreases) by 10\%.

Finally, we roughly estimate the present kinetic energy of the subcluster and the kinetic energy loss by the ICM heating and compare both.
The former, $K_{\rm s}$, can be estimated to be $(1/2) M_{\rm s} v_1^2$.
Here, $M_{\rm s}$ is the mass of the subcluster, which is obtained from the balance its self-gravity and pressure gradient,
i.e. $\sim kT_{\rm s} R_{\rm s} /\mu G m_{\rm H} \sim 3 \times 10^{13} M_{\odot}$,
where $T_{\rm s}$ is the ICM temperature, the $R_{\rm s}$ radius of the subcluster, 
$G$ the gravitational constant, and $m_{\rm H}$ the hydrogen mass.
Referring to the above discussions, $K_{\rm s}$ is estimated to be $\sim 10^{63}$ erg.
On the other hand, the energy loss by the ICM heating is estimated to be
$\Delta E_{\rm heat} = k \Delta T_{\rm heat} \times V_{\rm swept} \times \bar{n}$,
where $k \Delta T_{\rm heat}$ is the difference temperature between the preshock and the postshock region,
$V_{\rm swept} = \pi R_{\rm s}^2 D_{\rm tr}$ the total volume of the region swept by the subcluster,
and $\bar{n}$ the mean particle number density of the region.
$D_{\rm tr}$ is the distance along which the subcluster has traveled.
Using the values, $k \Delta T_{\rm heat} \lesssim 7.4$ keV, $R_{\rm s} = 150$ kcp, $D_{\rm tr} = 500$ kpc and $\bar{n} = 10^{-3}$ cm$^{-3}$,
we obtain $\Delta E_{\rm heat} \lesssim 8 \times 10^{60}$ erg.
The fact that $\Delta E_{\rm heat}$ is very much less than $K_{\rm s}$ means the subcluster is scarcely decelerated by ICM heating.

%%%%%% 5. Summary %%%%%%%%
\section{Summary}
The high-statistics map of the X-ray hardness ratio obtained from long-time ($\sim 100$ ks) observations of A 85
with Suzaku revealed a large peak on the east side of the subcluster.
From the spectral fittings, a comparable result was obtained.
A detailed temperature map shows a higher temperature ($\sim$ 8.5 keV) region that overlaps with this peak,
and a similar feature can be seen in the maps of pressure and entropy.
This high-temperature region (hot spot) has not been reported so far.
From the location of the hot spot, we consider that a postshock region produced by the infall of the subcluster from the southwest,
in contrast to previous studies, which have recognized that the subcluster is infalling from the southeast.

By using the Rankine--Hugoniot jump conditions for shocks,
Mach number and the infall velocity of the subcluster are estimated to be
$1.5 \pm 0.2$ and $1950^{+290}_{-280}$ km s$^{-1}$, respectively.
On the other hand, by using the redshifts of the subcluster's member galaxies obtained from optical observations,
the radial (line-of-sight) velocity difference between the subcluster and A 85 was found to be $520 \pm 240$ km s$^{-1}$,
in excess for the subcluster.
%spell ok
These velocities yield the conclusion that the angle between the line of sight and
the direction of the motion of the subcluster is $75^{+7}_{-8}$ degrees.
The present kinetic energy of the subcluster and the energy loss by the ICM heating
are roughly estimated to be $\sim 10^{63}$ and $\lesssim 8 \times 10^{60}$ erg, respectively.
This means that the subcluster is scarcely decelerated by ICM heating.

%%%%%% Acknowledgment %%%%%%%%
\bigskip
We thank an anonymous referee for useful comments.
Data analysis was carried out in part on the common-use data analysis computer system operated
the Astronomy Data Center (ADC) of the National Astronomical Observatory of Japan.
S.M. acknowledges support from a Kyoto Sangyo University Research Grant. 
We thank all the Suzaku team members for operating the satellite and assistance with data analysis.

%%%%% References %%%%%%

\end{document}